\documentclass[aps,prd,twocolumn,letterpaper,nofootinbib]{revtex4-2}

\usepackage{amsmath,amssymb}
\usepackage{geometry}
\begin{document}

\title{Absence of Quadratic-Order Sensitivity to Small Neutrino Mass Splittings in Disappearance Measurements}

\author{Sanjeev Kumar Verma}
\affiliation{Department of Physics and Astrophysics, University of Delhi, Delhi, INDIA - 110007}

\date{\today}

\begin{abstract}
Neutrino disappearance measurements using binned reconstructed-energy spectra exhibit a regime in which small mass-squared splittings become unidentifiable at quadratic order when smooth spectral shape uncertainties are represented by profiled nuisance parameters in the fit. In the small-phase limit, the oscillation-induced modification of the detected spectrum is quadratic in the mass-squared splitting and produces a smooth deformation of the reconstructed-energy distribution. If the nuisance deformation functions used in the fit can reproduce this energy dependence across the fitted bins, the quadratic oscillation-induced distortion can be absorbed by the systematic deformation space and the profiled chi-squared remains unchanged at this order. Sensitivity to the mass-squared splitting then arises only from higher-order oscillation effects or from restrictions imposed on the allowed smooth spectral freedom.
\end{abstract}

\maketitle

\section{Introduction}

Neutrino disappearance measurements probe neutrino mass-squared splittings through energy-dependent distortions of detected event spectra at fixed source--detector baselines \cite{Pontecorvo1968,GiuntiKim}. In practice, these measurements are performed using binned event rates as a function of a reconstructed observable, such as visible energy, charged-lepton energy, or recoil energy, rather than the neutrino energy itself. Sensitivity to oscillation parameters is extracted by fitting the reconstructed-energy spectrum, with correlated systematic uncertainties incorporated through nuisance parameters that allow smooth spectral deformations \cite{Fogli2002}. Spectral disappearance analyses of this type are standard in both long- and short-baseline experiments \cite{MINOS2013}.

A natural question therefore arises: what determines sensitivity to very small mass-squared splittings when the oscillation phase remains small across the energies contributing to the observed spectrum? In this regime, the leading oscillation-induced modification of the detected spectrum is smooth and quadratic in the mass-squared splitting, suggesting the possibility that it may be degenerate with smooth spectral deformations allowed by systematic uncertainties.

This work analyzes the origin of sensitivity to small neutrino mass-squared splittings in disappearance measurements. The focus is on the regime in which the oscillation phase remains small over the neutrino energies contributing to the reconstructed-energy bins used in the fit. In this limit, the survival probability admits a systematic expansion in powers of $\Delta m^2$, and the resulting modification of the reconstructed-energy spectrum takes a particularly simple form. The analysis shows that, once smooth energy-dependent systematic variations are allowed to vary freely, the leading oscillation-induced distortion becomes indistinguishable from an allowed systematic deformation. As a consequence, the profiled goodness of fit remains unchanged relative to the no-oscillation hypothesis at quadratic order in the mass-squared splitting.

The central result of this work is that, in the small-phase regime, the quadratic oscillation-induced spectral distortion becomes unidentifiable whenever it lies within the span of smooth deformation functions allowed by the nuisance-parameter model used in the fit. This reflects a general identifiability condition: an oscillation parameter cannot be constrained by spectral data when the leading oscillation-induced distortion lies within the space of spectral variations allowed by the systematic model. 

The result presented here rests on a small number of clearly defined assumptions. The analysis applies to disappearance measurements performed at fixed baseline using binned spectra in a reconstructed-energy observable, with sensitivity extracted through a chi-squared fit that profiles over nuisance parameters representing smooth, correlated spectral shape uncertainties \cite{Fogli2002}. The oscillation phase is assumed to remain small over the neutrino energies contributing to the fitted bins, so that the survival probability admits a systematic expansion in $\Delta m^2$. At quadratic order in this expansion, the oscillation-induced modification of the detected spectrum is smooth and non-oscillatory in reconstructed energy. Finally, it is assumed that the nuisance deformation functions used in the fit can represent this leading quadratic energy dependence. No assumptions are made about the microscopic origin of the detected spectrum, the detailed detector response, or the functional form of higher-order oscillation effects.

\section{Observed Spectrum and Fit Framework}

Detected events are binned according to a reconstructed-energy observable $E$. The data consist of observed event counts $N_i^{\rm obs}$ in bins labeled by index $i$, with representative reconstructed energies $E_i$. All event counts refer to detected final-state events and are functions of the reconstructed-energy observable.

The no-oscillation prediction $N_i^0$ denotes the expected event count in bin $i$ in the absence of oscillations. When oscillations are included but systematic deformations are not yet applied, the predicted spectrum is denoted by $N_i^{\rm osc}(\Delta m^2,\theta)$, where $\Delta m^2$ is the relevant neutrino mass-squared splitting and $\theta$ the associated mixing angle. In the limit of vanishing oscillations, this reduces to the no-oscillation prediction, $N_i^{\rm osc}(0,\theta)=N_i^0$.

Smooth energy-dependent systematic uncertainties are incorporated through nuisance parameters $\xi_k$ multiplying a set of smooth deformation functions $f_k(E)$ defined in reconstructed energy. These functions represent the correlated spectral shape variations allowed in the fit. The full predicted spectrum entering the statistical analysis is therefore written as \cite{Pontecorvo1968,GiuntiKim}

\begin{equation}
N_i(\Delta m^2,\theta,\boldsymbol{\xi})
=
N_i^{\rm osc}(\Delta m^2,\theta)
\left(1+\sum_k \xi_k f_k(E_i)\right).
\end{equation}

The agreement between prediction and data is quantified using a binned chi-squared statistic,

\begin{equation}
\chi^2
=
\sum_i
\frac{(N_i^{\rm obs}-N_i)^2}{\sigma_i^2}
+
\sum_k
\frac{\xi_k^2}{\sigma_{\xi_k}^2}.
\end{equation}

For each value of $\Delta m^2$, the chi-squared is minimized with respect to $\boldsymbol{\xi}$. The test statistic is

\begin{equation}
\Delta\chi^2(\Delta m^2)
=
\chi^2(\Delta m^2)-\chi^2_{\rm min}.
\end{equation}

\section{Small-Phase Expansion}

For neutrino disappearance at a fixed baseline $L$, the two-flavor survival probability as a function of the neutrino energy $E_\nu$ is \cite{Pontecorvo1968,GiuntiKim}

\begin{equation}
P_{\rm surv}(E_\nu;\Delta m^2,\theta)
=
1-\sin^22\theta\,\sin^2\!\left(\frac{\Delta m^2 L}{4E_\nu}\right).
\end{equation}

The oscillation phase is therefore

\begin{equation}
\phi(E_\nu)=\frac{\Delta m^2 L}{4E_\nu}.
\end{equation}

The analysis focuses on the small-phase regime defined by $|\phi(E_\nu)|\ll1$. This condition is understood to hold over the neutrino-energy range contributing significantly to each reconstructed-energy bin, so that the quadratic expansion remains valid after convolution with detector response and binning. In this limit,

\begin{equation}
\sin^2\phi(E_\nu)=\phi(E_\nu)^2+\mathcal{O}(\phi^4).
\end{equation}

We define the oscillation-induced change in the predicted number of detected events in reconstructed-energy bin $i$, prior to allowing any systematic deformations, as

\begin{equation}
\delta N_i \equiv N_i^{\rm osc}(\Delta m^2,\theta)-N_i^0.
\end{equation}

At quadratic order in $\Delta m^2$, the fractional distortion takes the generic form

\begin{equation}
\frac{\delta N_i}{N_i^0}
=
-\alpha(\Delta m^2,\theta)\,h(E_i)
+
\mathcal{O}\!\left((\Delta m^2)^4\right),
\end{equation}

where

\begin{equation}
\alpha(\Delta m^2,\theta)
=
\sin^22\theta
\left(\frac{L}{4}\right)^2
(\Delta m^2)^2 .
\end{equation}

The quadratic dependence arises from expanding the oscillation probability in the small-phase parameter inside the bin-level event-rate integrals that define the reconstructed-energy spectrum. The function $h(E_i)$ is defined by Eq.~(9) as the detector-level coefficient of $(\Delta m^2)^2$ in the fractional spectral distortion. Physically, $h(E_i)$ corresponds to a weighted average of the inverse-square neutrino-energy dependence over the neutrino energies contributing to reconstructed-energy bin $i$.

At this order, the distortion is non-oscillatory and varies smoothly with reconstructed energy. Smooth distortions of the neutrino-energy spectrum produce correspondingly smooth distortions of the reconstructed-energy spectrum after detector smearing. The quadratic oscillation-induced deformation therefore remains smooth in reconstructed energy.

\section{Degeneracy with Smooth Spectral Systematics}

Let $\{f_1,\ldots,f_K\}$ denote the set of smooth spectral deformation functions used in the fit. We assume that these deformation functions can approximate $h(E)$ at the level required by the quadratic expansion in $\Delta m^2$. That is,

\begin{equation}
h(E_i)=\sum_k a_k f_k(E_i)+r_i,
\end{equation}
where $r_i$ denotes a possible residual. In realistic analyses, individual nuisance modes such as overall normalization or a single energy-tilt deformation will not generally reproduce $h(E)$. The degeneracy becomes exact only when the combined space of allowed smooth shape variations is sufficiently flexible to approximate this dependence, in which case the residual $r_i$ controls the remaining sensitivity.

The first term represents the component of the oscillation-induced distortion contained within the nuisance deformation space, while the residual $r_i$ represents the component that cannot be reproduced by the systematic deformation model.

In typical spectral analyses, the allowed deformation functions represent smooth correlated variations of the predicted spectrum across reconstructed-energy bins. Over finite energy intervals, such smooth functions can approximate inverse-power energy dependences to the accuracy relevant at quadratic order in $\Delta m^2$. The residual $r_i$ therefore quantifies only the component of the quadratic oscillation distortion that lies outside the smooth deformation space defined by the systematic model.

The predicted event rates may then be written as

\begin{equation}
\begin{aligned}
N_i(\Delta m^2,\theta,\boldsymbol{\xi})
={}&
N_i^0
\Bigl[
1-\alpha h(E_i)+\sum_k \xi_k f_k(E_i)
\Bigr]
\\
&+\mathcal{O}((\Delta m^2)^4).
\end{aligned}
\end{equation}
Products such as $\alpha \xi_k$ contribute only at $\mathcal{O}((\Delta m^2)^4)$, since the canceling solution satisfies $\xi_k\sim\alpha$. Choosing $\xi_k=-\alpha a_k$ cancels the quadratic oscillation-induced distortion up to the residual $r_i$. Because the canceling nuisance shifts scale as $\xi_k \sim (\Delta m^2)^2$, the associated pull-term contribution $\xi_k^2/\sigma_{\xi_k}^2$ enters only at order $(\Delta m^2)^4$ for any finite prior width $\sigma_{\xi_k}$. Strong external constraints therefore modify the quartic sensitivity but do not restore quadratic-order identifiability unless the allowed deformation space itself is restricted.

\section{Implications for the Profiled Chi-Squared}

Because the chi-squared is minimized over all nuisance parameters, and because a canceling choice exists, the minimum chi-squared at fixed $\Delta m^2$ coincides with that obtained at $\Delta m^2=0$ up to corrections of order $(\Delta m^2)^4$ and any residual mismatch $r_i$. In the idealized case $r_i=0$,

\begin{equation}
\Delta\chi^2(\Delta m^2)=0
\qquad
\text{at } \mathcal{O}((\Delta m^2)^2).
\end{equation}

Sensitivity to $\Delta m^2$ in the small-phase regime therefore arises only from intrinsically higher-order oscillation effects or from restrictions imposed on the allowed smooth spectral deformations.

\section{Scope of Applicability}

The argument presented here applies generally to disappearance measurements in which the leading oscillation-induced modification of the detected spectrum is smooth and quadratic in the small expansion parameter. In three-flavor oscillations, disappearance probabilities contain sums of quadratic small-phase contributions proportional to $(\Delta m^2_{ij})^2$, and the resulting detector-level distortion remains smooth, so the same non-identifiability applies provided the nuisance deformation functions can represent the corresponding quadratic energy dependence. At quadratic order, these contributions differ only in their coefficients; they multiply the same inverse-square neutrino-energy dependence and therefore combine into a single smooth detector-level distortion. Consequently, the leading disappearance signal spans only one functional direction in reconstructed-energy space, so a single smooth deformation space can absorb the combined effect without introducing tension between different splittings.

This conclusion does not extend generically to appearance measurements. In disappearance channels, the oscillation probability is a sum of squared amplitudes, so that in the small-phase regime the leading modification of the detected spectrum is governed by a single smooth quadratic energy dependence. Appearance probabilities instead involve products of distinct oscillation amplitudes associated with different mass splittings and phases. When expanded in the small-phase regime, these terms generate multiple independent energy dependences, including contributions proportional to both $1/E_\nu$ and $1/E_\nu^2$. After detector convolution, the resulting spectral distortion generally spans more than one functional direction in reconstructed-energy space. Absorbing such distortions would require the systematic deformation space to reproduce multiple independent smooth modes simultaneously, which is not generically the case. Consequently, the quadratic-order cancellation identified for disappearance measurements does not extend in general to appearance channels.

\section{Conclusion}

In the small-phase regime of neutrino disappearance measurements, the leading oscillation-induced modification of reconstructed-energy spectra is quadratic in the mass-squared splitting and smooth in energy. When smooth spectral shape uncertainties are represented by profiled nuisance parameters and allowed to vary freely, this leading oscillation signal becomes degenerate with the systematic deformation functions used in the fit. As a result, the profiled chi-squared remains unchanged at quadratic order in the mass-squared splitting. Constraints obtained in this regime therefore probe either higher-order oscillation effects or modeling assumptions that restrict smooth spectral freedom, rather than the leading quadratic spectral deformation itself.

\section*{Data Availability Statement}

No new data were created or analyzed in this study.

\section*{Acknowledgments}

The author used AI-based tools to navigate the literature and assist with clarity of presentation. All scientific content, derivations, and conclusions were developed and verified by the author.

\end{document}